\documentclass{emulateapj}
\def\Hb{H\,{\sc $\beta$}}

\def\Civ{C\,{\sc iv}}
\def\Mgii{Mg\,{\sc ii}}
\def\Feii{Fe\,{\sc ii}}

\def\Siii{Si\,{\sc ii}}

\shorttitle{A ten billion solar mass black hole at $z=5.363$}
\shortauthors{Wang et al.}

\begin{document}

\title{An ultra-luminous quasar at \lowercase{$z=5.363$} with a ten billion solar mass black hole and a Metal-Rich DLA at  \lowercase{$z\sim5$}}

%% Begin Author
\author{Feige Wang\altaffilmark{1,2}, Xue-Bing Wu\altaffilmark{1,3}, Xiaohui Fan\altaffilmark{2,3}, Jinyi Yang\altaffilmark{1,2}, Zheng Cai\altaffilmark{2}, Weimin Yi\altaffilmark{4,5}, Wenwen Zuo\altaffilmark{6}, Ran Wang\altaffilmark{3}, Ian D. McGreer\altaffilmark{2}, Luis C. Ho\altaffilmark{1,3},Minjin Kim\altaffilmark{7}, Qian Yang\altaffilmark{1}, Fuyan Bian\altaffilmark{8,9}, Linhua Jiang\altaffilmark{3}}

\altaffiltext{1}{Department of Astronomy, School of Physics, Peking University, Beijing 100871, China}
\altaffiltext{2}{Steward Observatory, University of Arizona, 933 North Cherry Avenue, Tucson, AZ 85721, USA}
\altaffiltext{3}{Kavli Institute for Astronomy and Astrophysics, Peking University, Beijing 100871, China}
\altaffiltext{4}{Yunnan Observatories, Chinese Academy of Sciences, Kunming 650011,China}
\altaffiltext{5}{Key Laboratory for the Structure and Evolution of Celestial Objects, Chinese Academy of Sciences, Kunming 650011,China}
\altaffiltext{6}{Shanghai Astronomical Observatory, Chinese Academy of Sciences, Shanghai 200030, China}
\altaffiltext{7}{Korea Astronomy and Space Science Institute, Daejeon 305-348, Korea}
\altaffiltext{8}{Mount Stromlo Observatory, Research School of Astronomy and Astrophysics, Australian National University, Weston Creek, ACT 2611, Australia}
\altaffiltext{9}{Stromlo Fellow}

%% Begin paper

\begin{abstract}
We report the discovery of an ultra-luminous quasar J030642.51+185315.8 (hereafter J0306+1853) at redshift 5.363, which hosts a super-massive black hole (SMBH) with $M_{BH} = (1.07 \pm 0.27) \times10^{10}~M_\odot$. With an absolute magnitude $M_{1450}=-28.92$ and bolometric luminosity $L_{bol}\sim3.4\times10^{14} L_{\odot}$, J0306+1853 is one of the most luminous objects in the early Universe. It is not likely to be a beamed source based on its small flux variability, low radio loudness and normal broad emission lines. In addition, a $z=4.986$ Damped Ly$\alpha$ system (DLA) with $\rm [M/H]=-1.3\pm0.1$, among the most metal rich DLAs at $z \gtrsim 5$, is detected in the absorption spectrum of this quasar. This ultra-luminous quasar puts strong constraint on the bright-end of quasar luminosity function and massive-end of black hole mass function. It will provide a unique laboratory to the study of BH growth and the co-evolution between BH and host galaxy with multi-wavelength follow-up observations. The future high resolution spectra will give more insights to the DLA and other absorption systems along the line-of-sight of J0306+1853.
\end{abstract}


\section{Introduction}
Supermassive black holes (SMBHs) reside in the centers of most galaxies, with %typical 
masses in the range of $M_{BH}\sim 10^6-10^9M_{\odot}$. In the local universe, only a few objects have the measured black hole masses $M_{BH} \sim 10^{10}M_{\odot}$ \citep[e.g.][]{mcconnell11,vanden12}. Understanding when and how SMBHs form and grow is an important topic in extragalactic astronomy. One of the most direct ways to place constraints on the BH seeds and the growth of SMBHs is to study the masses of SMBHs hosted by luminous high redshift quasars \citep[e.g.][]{volonteri06,trakhtenbrot11}.

BH masses in quasars can be estimated by the single epoch virial mass estimate methods based on the low-ionization \Hb\ and \Mgii\ emission lines \citep[e.g.][]{vestergaard06,mclure04}. High-ionization lines such as \Civ\ can also be used \citep[e.g.][]{vestergaard06}; however, \Civ\ is thought to be less reliable than  \Hb\ and \Mgii\ in limited S/N and resolution data \citep[e.g.][]{shen11,trakhtenbrot12,denney13}.

The Sloan Digital Sky survey \citep[SDSS;][]{york00} and the SDSS-III Baryonic Oscillation Spectroscopic Survey \citep[BOSS;][]{dawson13} have provided the largest high redshift quasar sample and resulted in the discovery of several hundred $z\gtrsim4$ quasars with bolometric luminosity of several $10^{13} ~L_{\odot}$ \citep{schneider10,paris14}. The SDSS quasar sample has been used to study the quasar luminosity function \citep[QLF;][]{richards06a,ross13} and BH mass function \citep[BHMF;][]{kelly10,kelly13} at low and moderate redshifts. The BHMF suggested that  the maximum mass of a black hole in quasar to be $\sim 3\times10^{10}M_{\odot}$ \citep{kelly10,kelly13} at those redshifts. From the combination of SDSS quasars and Stripe 82 faint quasar sample, \cite{mcgreer13} presented the most detailed study on the $z \sim 5$ QLF, especially at the faint end. However, the slope of the QLF at the bright end still has large uncertainties \citep[e.g.][]{richards06a,mcgreer13}, mainly due to the small number of quasars observed to have extremely high luminosities (e.g. $M_{1450}\lesssim-27.5$).
%mainly due to less high luminosity quasars (e.g. $M_{1450}\lesssim-27.5$).

The discoveries of luminous $z\sim6$ quasars \citep[e.g.][]{fan01,willott10,banados14} and $z\gtrsim6.5$ quasars \citep[e.g.][]{mortlock11,venemans15} with luminosity of $\sim 10^{13}~L_{\odot}$ and \Mgii\ based BH mass of $\sim 10^{9}~M_{\odot}$, suggest that the fast growth of BH may start from very early epochs ($z\gtrsim20$) and involve massive seed BHs ($M_{seed}\gtrsim10^3M_{\odot}$) \citep[e.g.][]{volonteri06}. Recently, the first ten billion solar mass BH at $z>6$ has been discovered in an ultra-luminous quasar J0100+2802 at $z=6.30$, selected from SDSS and  Wide-field Infrared Survey Explorer \citep[WISE;][]{wright10} photometric data \citep{wu15}. This discovery posts significant challenge to the theory of BH formation and the early BH-galaxy coevolution again. 

\begin{figure*}
\centering
 \includegraphics[width=\textwidth]{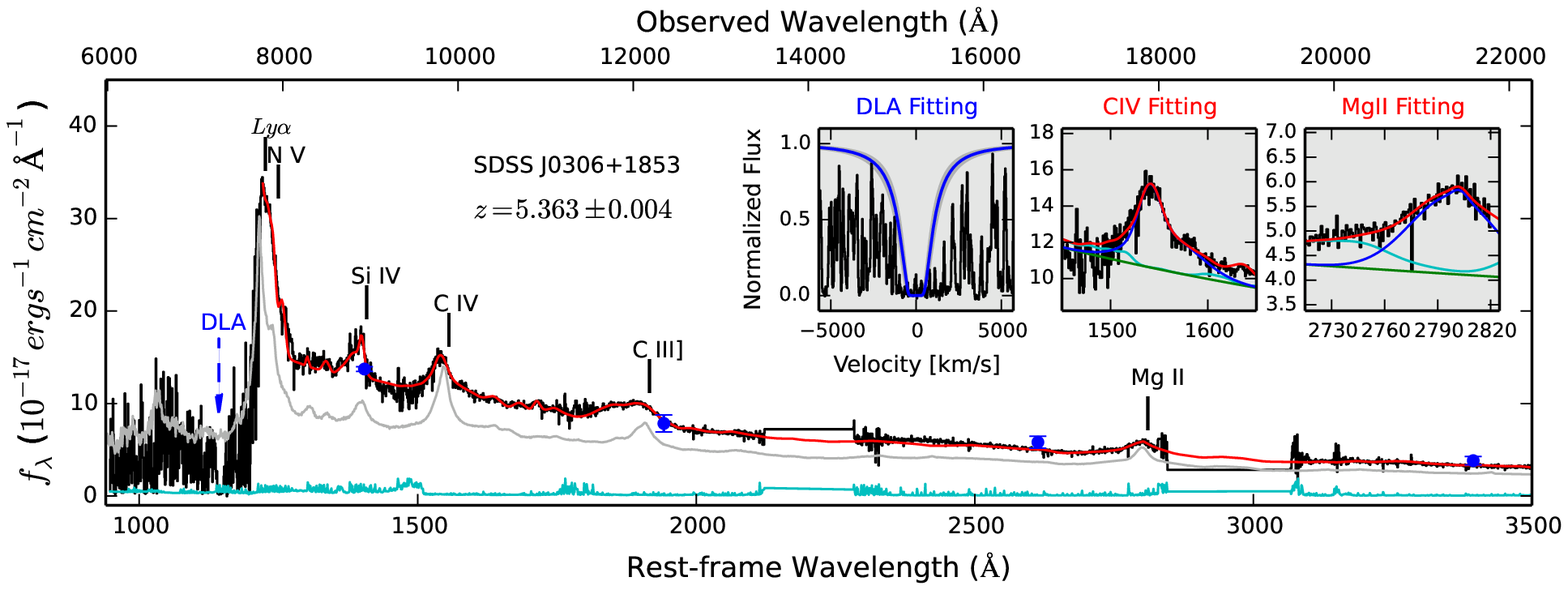}
  \caption{The spectrum of J0306+1853. The red line shows the best fitting and the grey line shows SDSS composite spectrum \citep{vanden01}. The cyan line shows the noise spectrum. The blue points, from left to right, represent SDSS $z$-band, 2MASS J, H and Ks photometry. The blue arrow labels the position of a DLA at $z=4.986$. The left inner plot shows the Voigt profile fitting of the DLA. Blue curve is the best-fit and gray shade includes 95\% limit. The column density of the DLA is $N_{\rm{HI}} = 10^{20.50^{+0.10}_{-0.12}}~\rm{cm^{-2}}$.
  Two right small insets show the fitting of \Civ\ and \Mgii\ emission lines; the green lines represent power-law and Balmer continuum, the cyan lines denote FeII contribution and the blue lines are contributions from  \Civ\ and \Mgii\ emissions. The gaps in the spectrum are due to the low sky transparency there.}
  \label{fig1}
\end{figure*}

WISE mapped the whole sky at 3.4, 4.6, 12, and 22 $\mu$m (W1, W2, W3, W4). \cite{wu12} suggested an efficient way to find high redshift quasars by combining SDSS and WISE photometric data, since most late-type stars have a significantly  bluer WISE colors than those of quasars. \cite{carnall15} used a similar method by combing optical and WISE colors to search $z\gtrsim5.7$ quasars within VST ATLAS survey. We are conducting a $z\gtrsim5$ quasar survey based on SDSS-WISE colors and have discovered several ultra-luminous quasars (F. Wang, et al. 2015, in preparation). Remarkably, we discovered an ultra-luminous quasar J0306+1853 at $z=5.363$ with a bolometric luminosity of $L_{bol}\sim3.4\times10^{14} L_{\odot}$ and a BH mass of $1.07\times10^{10}M_{\odot}$. 
J0306+1853,  along with J0100+2802,  are the only two quasars yet known with $M_{1450}\sim -29$ and the %Eddington based 
BH mass up to ten billion solar masses at $z>5$. Considering the high completeness of our selection method, these two quasars could be the only two ten billion solar mass BHs hosted by ultra-luminous quasars in entire the SDSS footprint.
We also present the detection of a  $z\sim5$ metally-enriched Damped Ly$\alpha$ system (DLA)  in the spectrum of J0306+1853.
Throughout this Letter, we use a $\Lambda$-dominated flat cosmology with $\rm H_{0}=70$ $\rm km s^{-1}Mpc^{-1}$, $\rm \Omega_{m}=0.3$ and $\rm \Omega_{\Lambda}=0.7$. The optical magnitudes in this letter are in AB system, and infrared magnitudes are in Vega system.

\section{Observations and Data}

J0306+1853 was selected as a high redshift quasar candidate based on SDSS and WISE photometric data (F. Wang, et al. 2015, in preparation). J0306+1853 is undetected in the u, g-bands but is relatively bright in the $r$, $i$, $z$ bands ($r=19.89\pm0.03$, $i=17.96\pm0.01$, $z=17.49\pm0.02$) and is strongly detected in four WISE bands ($\rm W1=14.31\pm0.03$, $\rm W2=13.46\pm0.04$, $\rm W3=10.51\pm0.10$, $\rm W4=7.59\pm0.16$). With $r-i=1.93$, $i-z=0.47$, J0306+1853 is in the area of the $griz$ color space strongly contaminated by M dwarfs \citep{richards02}. However, with $\rm W1-W2=0.86$, it is clearly separated from late-type stars. %using the combined SDSS+WISE photometry.  
J0306+1853 can also be selected by the z-W2/W1-W2 selection criteria used in \cite{carnall15}. Besides, J0306+1853 is also detected in Two Micron All Sky Survey \citep[2MASS;][]{skrutskie06} in J, H and Ks bands with $J=16.50\pm0.13$, $H=15.70\pm0.12$ and $Ks=15.12\pm0.13$, respectively. The overall shape of the spectral energy distribution (SED) of J0306+1853 built from SDSS, 2MASS and WISE photometry is consistent with the type I quasar composite SED \citep{richards06b}.
 
The first low-resolution optical spectrum of this source was obtained with the Lijiang 2.4m telescope using the Yunnan Fainter Object Spectrograph and Camera (YFOSC) on November 25, 2013 (UT). A 22-minutes exposure was taken with the G3 grism ($\lambda / \Delta \lambda \sim 680$ at 7000$\AA$) and a $1\farcs8$ slit. This low resolution spectrum clearly shows a sharp break at about 7700$\AA$, and is consistent with a quasar at redshift beyond 5.3. A possible $z\sim5$ DLA system is also visible in the spectrum. To confirm this, a higher resolution optical spectroscopic observation using Magellan Echellette spectrometer \citep[MagE;][]{marshall08} on the 6.5m Magellan/Clay Telescope, with 30 minutes integration time, was taken on January 3, 2014. Our configuration of the MagE optical spectrum provides a resolution of $\lambda / \Delta \lambda = 4100$ in the wavelength range from $6000\AA$ to 10,000$\AA$ with a $1\farcs0$ slit. The data were reduced by a custom-built pipeline (G. Becker, private communication) and flux calibrated using a standard star.  

In order to obtain the \Civ\ and \Mgii\ based BH masses and the metallicity of the DLA system in J0306+1853, we took a near-IR spectrum with the Folded-Port Infrared Echellette  \citep[FIRE;][]{simcoe10}, an IR echelle spectrograph on the 6.5m Magellan/Baade Telescope. The observation was carried out on January 17, 2014, with an integration time of 60 minutes, and provided a resolution of $R\sim5000$ from 0.82 to 2.5 $\mu m$. We reduced the data using the IDL based pipeline \citep[FIREHOSE;][]{simcoe11} and the flux was calibrated with a nearby A0V telluric star. The magnitudes estimated from MagE and FIRE spectra are consistent with the SDSS and 2MASS photometric data ($\Delta m \lesssim 0.2$) and thus imply a small variability of J0306+1853 in a long time baseline. Finally, we combined the MagE optical spectrum and FIRE infrared spectrum and scaled it to match the SDSS $z$-band and 2MASS J, H and Ks-bands photometric data for the absolute flux calibration (See Figure 1.). The final spectrum yields a redshift of $z=5.363\pm 0.004$ based on the \Mgii\ line fitting.

J0306+1853 is undetected in the NRAO VLA Sky Survey \citep[NVSS;][]{condon98}, the completeness limit of which is about 2.5 mJy at 1.4 GHz. This constrains the radio loudness of J0306+1853 to be $R = f_{6cm} /f_{2500}\lesssim76$ following \cite{shen11}, where $f_{6cm}$ and $f_{2500}$ are the flux density ($f_\nu$) at rest-frame 6 cm and 2500 \AA, respectively. The radio loudness upper limit together with the small variability suggest J0306+1853 is not a beamed source.

\section{Luminosity and black hole mass measurements}
\begin{figure}
\centering
 \includegraphics[width=0.5\textwidth]{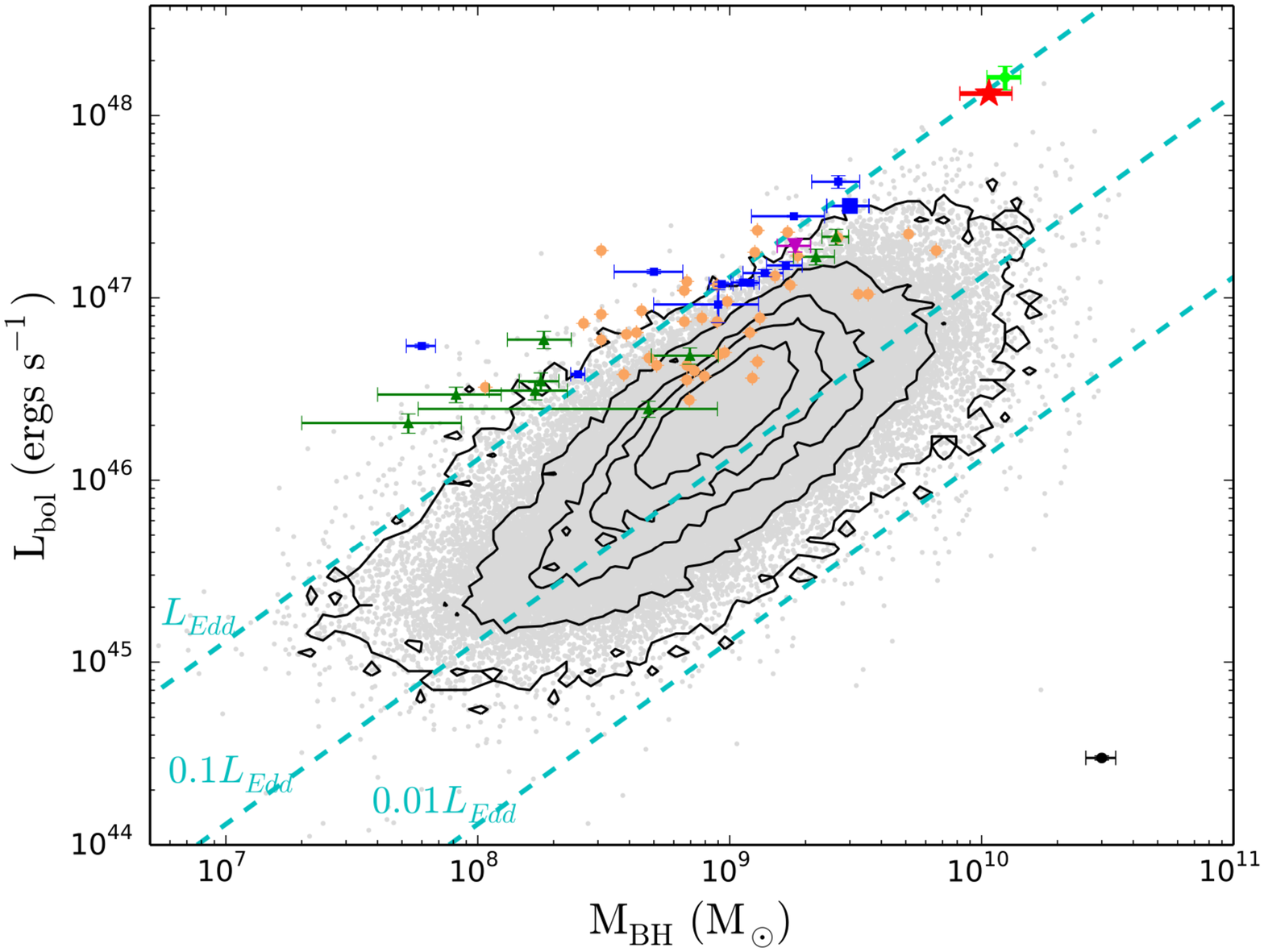}
  \caption{Distribution of quasar bolometric luminosities and black hole masses estimated from \Mgii\ lines. The Red star represents J0306+1853 and the light green cycle represents the luminous quasar J0100+2802 at $z=6.30$ \citep{wu15}. The brown points denote $z\sim4.8$ quasars from \cite{trakhtenbrot11}. The blue squares denote SDSS high redshift quasars at $z\sim 6$ \citep{kurk07,jiang07,derosa11}. The green upper triangles denote CFHQS high-redshift quasars at $z\sim 6$ \citep{willott10,derosa11}. The purple lower triangle denotes the most distance quasar ULAS J1120+0641 at z=7.085 \citep{mortlock11}. Black contours and grey dots denote SDSS low redshift quasars from \cite{shen11} (with broad absorption line quasars excluded). The error bars represent the 1$\sigma$ standard measurement errors, and the mean error bar for low redshift quasars is presented in the bottom-right corner. The dashed lines denote the luminosity in different fraction of Eddington luminosity. Note that the systematic uncertainties (not included in error bars) of virial BH masses could be up to a factor of 0.5 dex.}
  \label{fig2}
\end{figure}

To derive line widths and continuum luminosities used for single-epoch virial BH mass estimators, we performed least-$\chi^2$ global fitting to the combined optical and near-IR spectrum. The spectrum was de-reddened for Galactic extinction using the \cite{cardelli89} Milky Way reddening law and E(B-V) derived from the \cite{schlegel98} dust map. The spectrum was then shifted to rest frame using the redshift based on \Mgii\ . We first fit a pseudo-continuum model to account for the power-law continuum, \Feii\ emission, and Balmer continuum in the line free region. Templates of \Feii\  emission were from \cite{shen11} by combining the \cite{vestergaard01} template (1000-2200$\AA$) and the \cite{salviander07} template (2200-3090$\AA$). The \Feii\  template is broadened by convolving the template with Gaussians with a range of intrinsic line widths and then scaled to match the observed spectrum. We model the Balmer continuum following \cite{derosa11}, which assumed a partially optically thick gas clouds with uniform temperature $T_{e}=15,000K$ and normalized to the power-law continuum at $\lambda_{rest}=3675\AA$. 
From the best pseudo-continuum fit, we measured the rest-frame $\rm 3000~\AA$ and $\rm 1350~\AA$ power-law luminosities as $\rm 2.04\times10^{47}~ergs~s^{-1}$ and $\rm 3.46\times10^{47}~ergs~s^{-1}$, and the absolute magnitude at rest-frame $\rm 1450~\AA$ as $\rm M_{1450,AB}=-28.92\pm0.04$. By assuming an empirical conversion factor from the luminosity at $\rm 1350~\AA$ \citep{shen11}, we estimate the bolometric luminosity as $L_{bol} = 3.81\times L_{\rm{1350~\text{\normalfont\AA}}} = (1.32\pm0.08) \times 10^{48}~\rm{ergs~s^{-1}}$. We get the Eddington mass to be $\sim10^{10}M_{\odot}$ by assuming Eddington accretion, which is an approximate lower limit of the BH mass \citep[e.g.][]{kurk07}.

We then fit individual emission lines after subtracting the pseudo-continuum.  As two gaussian model provided a sufficiently good model of the line profiles of \Mgii\ and \Civ\ given the S/N of our spectra, we use two Gaussians to fit \Mgii\ and \Civ\ emission lines, respectively. For other emission lines, we use one Gaussian to fit isolated lines and multiple Gaussians to fit blended lines, same as did in \cite{jiang07}. The best fitting is  shown in Figure 1. We noticed that a portion of red part of \Mgii\ profile is missed in the FIRE spectrum due to the low transmission, which may cause some uncertainties in the estimation of Mg II line width. The FWHMs and equivalent widths (EWs) of \Mgii\ and \Civ\ are $\rm 5722 \pm 640~km~s^{-1}$, $22.6\pm2.8\AA$ and $\rm 7320 \pm 680~km~s^{-1}$, $25.1\pm2.1 \AA$, respectively. 

To estimate the black hole mass of J0306+1853, we adopt the relation obtained by \cite{vestergaard09} for \Mgii\ estimator:
\begin{equation}
\small
\frac{M_{BH}\rm{(Mg \ II)}}{M_\odot} = 10^{6.86} \left[\frac{\lambda L_{\lambda}\rm{(3000~\AA)}}{\rm{10^{44}~ ergs~s^{-1}}}\right]^{0.5}  \left[\frac{\rm{FWHM(Mg \ II)}}{\rm{10^3~km~s^{-1}}}\right]^2
\end{equation}

and \cite{vestergaard06} for \Civ\ estimator:

\begin{equation}
\small
\frac{M_{BH}\rm{(C \ IV)}}{M_\odot}=10^{6.66} \left[\frac{\lambda L_{\lambda}\rm{(1350~\AA)}}{\rm{10^{44}~ ergs~s^{-1}}}\right]^{0.53}  \left[\frac{\rm{FWHM(C \ IV)}}{\rm{10^3~km~s^{-1}}}\right]^2
\end{equation}

These relations give the black hole mass of J0306+1853 as ${M_{BH}\rm{(Mg \ II)}}=(1.07 \pm 0.27) \times10^{10}~M_\odot$ and ${M_{BH}\rm{(C \ IV)}}=(2.16 \pm 0.48) \times10^{10}~M_\odot$. The Eddington ratio is $L_{bol} / L_{Edd} = 0.95$ for \Mgii\ based measurement and 0.47 for \Civ\ based measurement. The uncertainties of BH mass is measured from the 68\% range of the distributions from fitting 200 mock spectra generated using the method in \cite{shen11}. Note that it does not include systematic errors of using different BH mass estimators. By using different FeII templates \citep[e.g.][]{salviander07,tsuzuki06} and different estimators \citep[e.g.][]{mclure04,vestergaard09,wang09,shen11,trakhtenbrot12}, the \Mgii\ based BH mass of J0306+1853 could be in the range of 7.3-20.7 billion solar mass and the systematic uncertainties of MgII based BH masses could be $\sim$0.5 dex \citep[e.g.][]{shen13}. However, the \Mgii\ based BH mass measurement is still the most reliable method for measuring BH masses at high redshift. We will only use the \Mgii\ based BH mass for subsequent discussion. Figure 2 shows the distribution of quasar bolometric luminosities and black hole masses estimated from the \Mgii\ lines for a large sample of quasars from literatures over a wide range of redshifts. Adopting the $M_{BH}$ estimate based on \cite{mclure04} calibration, the BH in J0306+1853 is more massive by $\sim0.25$ dex than the most massive one found in \cite{trakhtenbrot11}, or by 0.38 dex compared with the \cite{derosa11} study. It is close to the most massive black holes at any redshift, and together with J0100+2822 \citep{wu15}, the most massive BH yet discovered at $z>5$. Considering the high completeness of our selection method, J0306+1853 and J0100+2802 could be the only two ten billions solar mass BHs hosted by ultra-luminous quasars at $z>5$ in the SDSS footprint.

\section{A Damped $L \lowercase{y \alpha$} System at \lowercase{$z\sim5.0$}}

\begin{figure}
\centering
\includegraphics[width=0.5\textwidth]{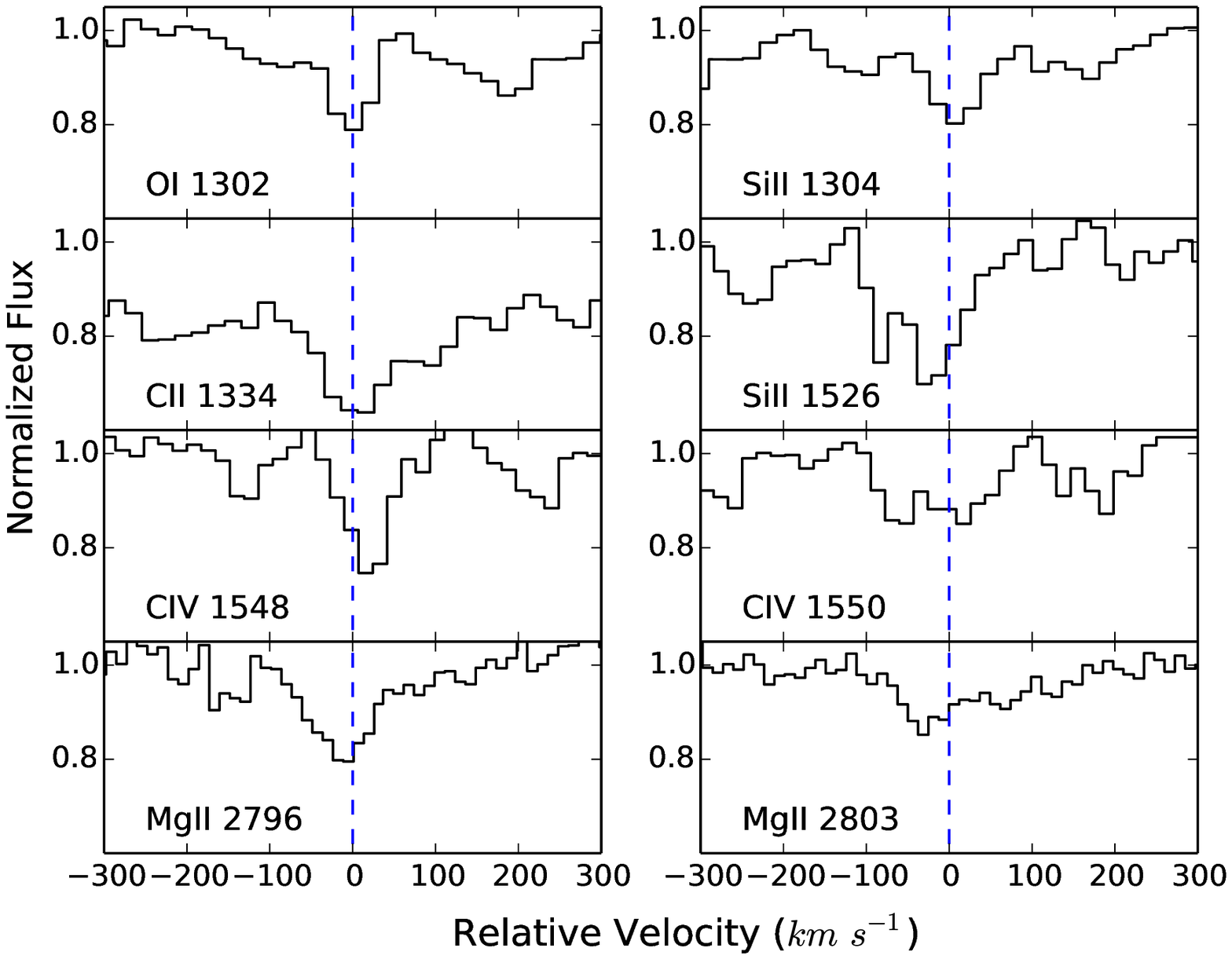}
  \caption{Velocity profiles of metal transitions for the $z=4.986$ DLA. The total column density of Si is $N_{\rm{Si}}=10^{14.8}~\rm{cm^{-2}}$ and the metallicity of this DLA is [M/H] $=-1.3\pm 0.1$. See \S4 for details.}
 \label{fig3}
\end{figure}

Damped Ly$\alpha$ systems (DLAs) are atomic hydrogen gas clouds measured in absorptions to background quasars with a column density higher than $2\times 10^{20}$ $\rm cm^{-2}$, which are unique laboratories for understanding the conversion of neutral gas into stars at high redshift \citep{wolfe05}. However, the number of known high redshift DLAs are very rare 
and only about ten DLAs have been discovered at $z>4.7$ \citep{rafelski12}. The studies of metallicities of these high redshift DLAs suggest a rapid decline in metallicity of DLAs at $z\sim5$ \citep{rafelski12,rafelski14}.

A $z=4.986$ DLA system clearly presents in the absorption spectrum of J0306+1853 (Figure 1.). We determine the $N_{\rm {HI}}$ value of the DLA system by fitting a Voigt profile using the method of \cite{rafelski12}. The best fit yields a column density of $N_{\rm{HI}} = 10^{20.50^{+0.10}_{-0.12}}~\rm{cm^{-2}}$. We estimate the error on the $N_{\rm {HI}}$ value by manually selecting values that ensure that $N_{\rm {HI}}$ solutions with assuming a single component of DLA plus possible deblending $Ly\alpha$ forest in the wings. 
%However, we need to caution that the systematic effects of line saturation and continuum placement will lend larger uncertainties in most cases \citep[e.g.][]{rafelski12}. 
%According to \cite{rafelski12}, our one-sigma error should be regarded as 0.1 dex systematic uncertainties for all the measurements. 
However, we need to caution that the systematic effects of line saturation and continuum placement will lend larger uncertainties in most cases \citep[e.g.][]{rafelski12}, and the systematic uncertainties for all the measurements could be $\sim 0.1$ dex.

This DLA is also associated with a number of corresponding metal lines,  including {C\,{\sc ii}} $\lambda$1334, \Siii\ $\lambda$1304, $\lambda$1526, {O\,{\sc i}} $\lambda$1302, \Civ\ $\lambda$1548, $\lambda$1550 and \Mgii $\lambda$2796, $\lambda$2803 (Figure 3). \cite{rafelski12} suggests that S and Si are ideal elements to determine metallicity. We use the apparent optical depth method \citep[AODM;][]{savage91} to derive the total column density of Si with \Siii\ $\lambda$1304, $\lambda$1526 absorption lines, and find $N_{\rm{Si}}=10^{14.8}~\rm{cm^{-2}}$. 
The rest-frame equivalent width of \Siii\ $\lambda$1304 and \Siii\ $\lambda$1526 are $0.070\pm0.005\AA$ and $0.10\pm0.007\AA$, respectively. 
We estimate the metallicity of this DLA is [M/H] $=-1.3\pm 0.1$. Note that the error does not include systematic uncertainties. This makes it among the most metal-rich DLAs found to date at $z\gtrsim5$. Although the metallicity of the DLA in J0306+1853 is about 0.7 dex higher than the average metallicity of other $z\sim5$ DLAs, the existence of such system is still consistent with the rapid decline in metallicity of DLAs at $z\gtrsim5$  \citep{rafelski14}.

\section{Discussion and Summary}
Although gravitational lensing is a possible explanation for its high luminosity and inferred BH mass, we do not expect a large lensing magnification as J0306+1853 is an unresolved object in the FIRE acquisition image under the $1\farcs0$ seeing condition. Besides, J0306+1853 is not possible to be a beamed source as implicated from the small flux variability, small radio loudness and the normal broad emission lines. The bright-end slope of $z\sim5$ QLF is measured to be $\beta \lesssim -4$, indicating a rapid decline in the space density of luminous quasars at high redshift \citep{mcgreer13}. We searched the entire SDSS footprint ($\sim 14,000$ deg$^2$) with our SDSS-WISE selection method and find J0306+1853 to be the only quasar with such high luminosity at this redshift. Since our selection method is highly complete for luminous quasars at $z\gtrsim5$ (J. Yang, et al. 2015, in preparation), J0306+1853 is possible to be the only quasar with $M_{1450}<-28.9$ at redshift $z\gtrsim5$ in the SDSS footprint. From the prediction using the $z\sim5$ QLF \citep{mcgreer13}, the number of quasars with such high luminosity at $4.7\lesssim z \lesssim 5.4$ is about two in the whole sky and the probability to find one such luminous quasar in the SDSS footprint with our selection method is $\sim$35\%. 
Together with J0100+2802 at higher redshift, J0306+1853 is the only other known quasar with $M_{1450}$ up to $\sim-29$ at $z>5$. Although the discovery of J0306+1853 is consistent with the expectation from the current measurements of the QLF, its existence (and our finding of only a single object at this luminosity) shows that the luminosity function extends to much greater luminosities than previously explored, and sets a strong constraint on the bright end slope. A detailed studies of the QLF will be presented in a forthcoming paper (J. Yang, et al. 2015, in preparation) based on a complete sample of SDSS$-$WISE selected $z\sim5$ quasars.

\cite{kelly13} estimated the BHMF and Eddington ratio function (BHERF) at $z\lesssim5$ for Type 1 quasars from uniformly selected $\sim58,000$ SDSS quasars. Assuming there is no evolution from redshift of 4.75 to 5.25, their BHMF predicts that there should be several tens quasars with BH mass large than 10 billion $M_{\odot}$ at redshift between 5.0 and 5.5. 
Since J0306+1853 is the only quasar discovered in our survey with $M_{BH}$ of $10^{10}$ $M_\odot$ and Eddington ratio close to the order unity discovered, this would imply that there is a population of fainter quasars with lower Eddington ratio hosting $10^{10}$ $M_\odot$ black holes. A complete survey of such objects requires high quality near-IR spectroscopy of a large sampe of fainter $z\sim5$ quasars.

Absorption spectra of high-redshift quasars provides one of the most powerful tools to study intergalactic medium (IGM) in the early universe \citep[e.g.][]{simcoe11}. Future high signal-to-noise, high resolution spectra of ultra-luminous quasars as J0306+1853 will be very promising probes to the study of the IGM evolution and the shape of the ionizing background at $z\gtrsim5$.

\acknowledgments
We thank the referee for providing constructive comments and suggestions.
F. Wang and X.-B. Wu thank the supports by the NSFC grant No.11373008, the Strategic Priority Research Program ''The Emergence of Cosmological Structures'' of the Chinese Academy of Sciences, Grant No. XDB09000000, and the National Key Basic Research Program of China 2014CB845700. X. Fan and I. D. McGreer thank the support from the US NSF grant AST 11-07682. Funding for the Lijiang 2.4m telescope is provided by Chinese Academy of Sciences and the People's Government of Yunnan Province. This paper includes data gathered with the 6.5 meter Magellan Telescopes located at Las Campanas Observatory, Chile. We thank George Becker for providing MagE spectral data reduction. This research uses data obtained through the Telescope Access Program (TAP), which has been funded by the Strategic Priority Research Program "The Emergence of Cosmological Structures" (Grant No. XDB09000000), National Astronomical Observatories, Chinese Academy of Sciences, and the Special Fund for Astronomy from the Ministry of Finance.We acknowledge the use of SDSS, 2MASS, WISE photometric data. 

{\it Facilities:}  \facility{2.4m/YNAO (YFOSC)}, \facility{Magellan (MagE)}, \facility{Magellan (FIRE)}, \facility{Sloan (SDSS)}, \facility{WISE}, \facility{2MASS}.

\clearpage

\end{document}